\pdfoutput=1
\documentclass[]{aastex63}

\received{July 7, 2020}
\revised{September 1, 2020}
\accepted{September 15, 2020}
\submitjournal{PSJ}

\shorttitle{Convex shape of 11351 Leucus}
\shortauthors{Mottola et al.}

\begin{document}

\title{Convex Shape and Rotation Model of Lucy Target 
(11351) Leucus\\ from Lightcurves and Occultations}

\correspondingauthor{Stefano Mottola}
\email{stefano.mottola@dlr.de}

\author[0000-0002-0457-3872]{Stefano Mottola}
\affiliation{Institute of Planetary Research, DLR \\
Rutherfordstr. 2, \\
12489 Berlin, Germany}

\author{Stephan Hellmich}
\affiliation{Institute of Planetary Research, DLR \\
Rutherfordstr. 2, \\
12489 Berlin, Germany}

\author[0000-0003-0854-745X]{Marc W. Buie}
\affiliation{Southwest Research Institute \\
1050 Walnut St. \\
Boulder, CO 80302 USA}

\author{Amanda M. Zangari}
\altaffiliation{Amanda Zangari is currently an MIT Lincoln Laboratory employee. 
No Laboratory funding or resources were used to produce the results/findings reported in this publication.} 
\affiliation{Southwest Research Institute \\
1050 Walnut St. \\
Boulder, CO 80302 USA}

\author{Simone Marchi}
\affiliation{Southwest Research Institute \\
1050 Walnut St. \\
Boulder, CO 80302 USA}

\author{Michael E. Brown}
\affiliation{California Institute of Technology \\
1200 E. California Blvd. \\
Pasadena, CA 91125 USA}

\author{Harold F. Levison}
\affiliation{Southwest Research Institute \\
1050 Walnut St. \\
Boulder, CO 80302 USA}



\begin{abstract}

We report new photometric lightcurve observations of the Lucy Mission target (11351) Leucus
acquired during the 2017, 2018 and 2019 apparitions. We use these data in combination 
with stellar occultations captured during five epochs \citep{Buieetal2020} to 
determine the sidereal rotation period, the spin axis orientation,  a convex shape model, the absolute scale of the object, its geometric albedo, 
and a model of the photometric properties of the target. We find that Leucus is a prograde rotator with
a spin axis located within a sky-projected radius of 3$\degr$ (1$\sigma$) from J2000 Ecliptic coordinates ($\lambda=208\degr$, $\beta=+77\degr$) 
or J2000 Equatorial Coordinates (RA=248$\degr$, Dec=+58$\degr$). The sidereal period is refined to $P_{sid}=445.683\pm0.007$ h.
The convex shape model is irregular, with maximum dimensions of (60.8, 39.1, 27.8) km. The convex model accounts for 
global features of the occultation silhouettes, although minor deviations suggest that local and global concavities are present.
We determine a geometric albedo $p_V=0.043\pm0.002$.
\added{The derived phase curve supports a D-type classification for Leucus}.

\end{abstract}



\section{Introduction} \label{sec:intro}

Jupiter Trojans are a class of small objects trapped in the
Jupiter L4 and L5 Lagrangian points. Their origin is still disputed,
with the most likely scenarios falling into the categories 1) coeval trapping of local planetesimals in the 1:1 mean motion resonance with 
accreting Jupiter as a consequence of drag or collisions \citep{Yoder1979,Shoemakeretal1989} 
or 2) post Jupiter-formation 
capture of scattered trans-Neptunian planetesimals following an episode
of orbital chaos \citep{Morbidellietal2005} during the orbital migration of the giant planets or 
as a consequence of the \textit{Jumping Jupiter} scenario as defined in \citet{Nesvornyetal2013}. In either cases,
Trojans are thought to be primitive objects that experienced little thermal evolution
and contain a considerable amount of volatiles, which makes them 
close relatives to cometary nuclei. \\

With a launch planned for October 2021, Lucy is the thirteenth NASA mission of 
the Discovery Program and will be the first one to explore the Jupiter Trojan System. 
Its trajectory is designed to fly-by 5 Trojans -- one of which, (617) Patroclus is an equal-size binary system --
distributed over the two Lagrangian clouds. The encounter with Leucus is currently planned for June 18, 2028.
A coordinated effort has been initiated to support the mission with 
a systematic program of ground-based observations of the mission targets
aiming at characterizing their dynamical, physical, rotational and 
photometric properties. The goal is to inform the mission design in 
order to maximize the scientific return of the encounters and to complement
the space-based measurements with data that are most efficiently 
acquired from the ground. \\

This paper presents new lightcurve photometry of the Lucy target 
Leucus, which is used, together with the results of stellar occultation 
campaigns presented in a companion paper \citep{Buieetal2020},
to determine its convex shape, spin axis orientation, albedo, size, sphere-integrated phase curve
and V--R color index.

\section{Leucus} \label{sec:leucus}

(11351) Leucus belongs to the Jupiter L4 Trojan swarm. Radiometric measurements by 
the IRAS satellite reported a size and a geometric albedo of 42.2$\pm{4.0}$ km and 0.063$\pm{0.014}$, respectively \citep{Tedescoetal2004}. 
\citet{Gravetal2012}, reported size and albedo of 34.155$\pm{0.646}$ km and 0.079$\pm{0.013}$, respectively, based on WISE radiometry.
\citet{Levison2016} classified Leucus tentatively as belonging to the D taxonomic type, \added{based on its visible spectral slope \citep{Roigetal2008}}.
No collisional family membership has been proposed as of today. 
\citet{Frenchetal2013} first realized from lightcurve data that Leucus is a very slow rotator, 
although they identified an incorrect period of 515 h. \citet{Buieetal2018}, by using 
photometric observations acquired during the 2016 apparition, combined with the 
observations by French, determined a firm rotation period of 445.732 $\pm$ 0.021 h.
They also estimated a geometric albedo of 0.047 based on the WISE diameter and on 
an average color index for D-type asteroids.

\section{Observations and data reduction} \label{sec:observations}
The new Leucus photometric observations reported in this paper were performed during its 2017, 2018 and 2019 apparitions 
by using 1.0 m telescopes from the Las Cumbres Observatory Global Telescope (LCOGT) network,
the 1.2 m telescope at the Calar Alto Observatory, Spain, and the two 24$''$ telescopes sited at Sierra Remote Observatories (SRO), Auberry, CA, USA, owned and operated by SwRI.
The observational circumstances are detailed in Table  \ref{tab:obscirc}. 
Typical exposure times were of 5 min for the Calar Alto observations, with the 
telescope tracked at half the relative tracking rate of the asteroid, in order reduce smearing and
obtain equal point-spread functions for the target and field stars. The LCOGT observations were also exposed for typically 5 min, 
but were tracked at object rate, since those telescopes do not support halfway tracking.

The SRO systems use an Andor Xyla sCMOS camera with a maximum exposure time of 30 seconds.  
All data taken on Leucus used this maximum exposure time.  In 30 seconds, the motion of Leucus is 125 mas at opposition, 
corresponding to less than a half of a pixel smear and and even smaller fraction of the point-spread function (PSF).  
The very fast readout and low read noise of the sCMOS camera nearly eliminate the penalties of taking such 
short exposures.  During processing, we can stack as many images as desired to reach a SNR goal.  
Two stacks are built from the data, one registered on the stars and the other registered on the apparent 
motion of Leucus.  We can use image subtraction to remove the stars in the Leucus stack but this was 
not necessary for the 2018 and 2019 data from SRO\null.  For these data we chose to stack 10 images 
at a time and used synthetic aperture integration to retrieve the photometry of Leucus, thus providing 
an effective integration over 5 minutes.  The star-stacked images were used with the same aperture 
to determine the instrumental magnitudes of the stars.

This raw photometry was further binned in time by a factor of 3 to increase the 
per-point SNR while also allowing the estimation of a good uncertainty for the photometry.

For the observations from Calar Alto, 
most fields were measured with \textit{Himmelspolizey}, a reduction pipeline developed by SH. The latter implements a semi-automatic 
astrometric/photometric workflow that uses \textit{SExtractor} \citep{Sextractor} for photometric extraction, an 
optimistic pattern matching algorithm \citep{Tabur2007} for astrometric 
reduction and a moving object detection algorithm for asteroid identification 
as described in \citet{Kubicaetal2007}.
In the case of
crowded fields, synthetic aperture photometry was measured interactively with AstPhot \citep{Mottolaetal1995}.\\
  
The observations proved to be challenging mainly because of two aspects. 
First, during the 2017 apparition the target was still close to the Galactic center, which
implied that involvement of the source with field stars was extremely frequent.
This issue was dealt with by applying star subtraction. For the observations
taken from the LCOGT Network the subtraction was performed 
in the image domain, by applying the method described in \citet{Buieetal2018}.
For the observations from Calar Alto, the subtraction was performed in the 
intensity domain, by separately measuring the flux of the involved stars at many epochs distant from the 
time of the respective appulses. Although star subtraction mitigates the contamination problem,
it is not a perfect solution. While the flux of the polluting star is removed, the photon noise associated to the 
subtracted flux still contributes to the degradation of the total SNR, sometimes being the dominant contribution. 
In the latter cases the affected frames were excluded altogether from the data set.
Furthermore, imperfect background source removal, due e.g. to changing seeing conditions
during the observations, results in a non-Gauss distribution of the measurements error which can cause outliers in the data
that can be difficult to discern.\\

The second challenging aspect was represented by the very long rotation period of the target,
which caused a useful night of observations to produce data points only at a single rotational aspect.
As a consequence, a large number of nights over extended periods of time were needed to 
ensure complete rotational coverage. Furthermore, the resulting data set virtually consisted only of
sparse photometry, in the sense that subsequent nights' observations didn't overlap in rotation phase,
preventing a composite lightcurve to be compiled from relative photometry. Composite lightcurves
were therefore generally constructed based on absolute photometry, with the disadvantage of 
the additional uncertainty component due to the errors of the calibration zero points.
Fortunately, modern, all-sky photometric catalogs with very good coverage, limiting magnitude and 
accuracy have become available in recent times, so that the problem of the accuracy of the zero point is not as severe as in the past.
With the fields of view of our detectors, a large number of suitable photometric catalog stars was always simultaneously present in the same frame as the target,
allowing us to identify variable stars and outliers.
For the observations from LCOGT, carried out with the SDSS r' filter,  we used the APASS photometric catalog \citep{Hendenetal2012}.
For the Calar Alto observations, carried out in the Johnson V and Cousins R$_C$ filters, we used the GAIA DR2 catalog \added{\citep{GaiaCollaboration2018}} 
with the transformations from the G photometric band from \citet{Evansetal2018}.
The observations from the SRO were performed with a VR broadband filter.
Gaia DR2 field stars were used to directly express the asteroid 
in the Gaia G bandpass without further transformation or color correction. 

Typically, relative photometric accuracy of the individual points (binned points for the SRO) ranged from
0.01 to 0.03 mag RMS. The absolute photometric accuracy of the zero points was typically of the order of 0.02 mag RMS. 
\added{For all observations, the magnitudes were reduced to 1 au from the Sun and the observer.
The times were converted to the Barycentric Dynamical Time (TDB) frame, in order to provide a uniform time reference.
Further, the times were corrected for the one-leg, target-observer light-travel time.\\
 }

For the purpose of compact representation -- but not for model computation --  the photometric time series were compiled into composites for each individual apparition.
This was achieved by performing a Fourier series fit of the fourth order to determine the respective best-fit synodic periods
by using the procedure described in \citet{Harrisetal1989}. Normally, we would fit simultaneously the Fourier coefficients and
a phase function to absolute photometry data, in order to compensate for brightness changes due to the phase curve. The implicit assumption
is that the shape of the lightcurve -- in particular its amplitude -- remains constant over a few consecutive rotational cycles. While this is a reasonable assumption for most asteroids, 
in the case of Leucus, given its 
very slow rotation, the amplitude of the lightcurve can change over consecutive rotational cycles due to the change in viewing and observing geometry.
Fitting the phase function simultaneously with the rotation period would tend to compensate the change in the lightcurve amplitude 
by skewing the phase function, which could result in varying phase curve slopes for different apparitions. 
For this reason we performed the composites with a nominal linear phase coefficient $\beta=0.0395$ $mag/\degr$ for all 
of the observations reported in this paper \added {(see section \ref{subsec:phasecurve})}.\\

\begin{figure}
\centering
  \includegraphics[width=13.8cm]{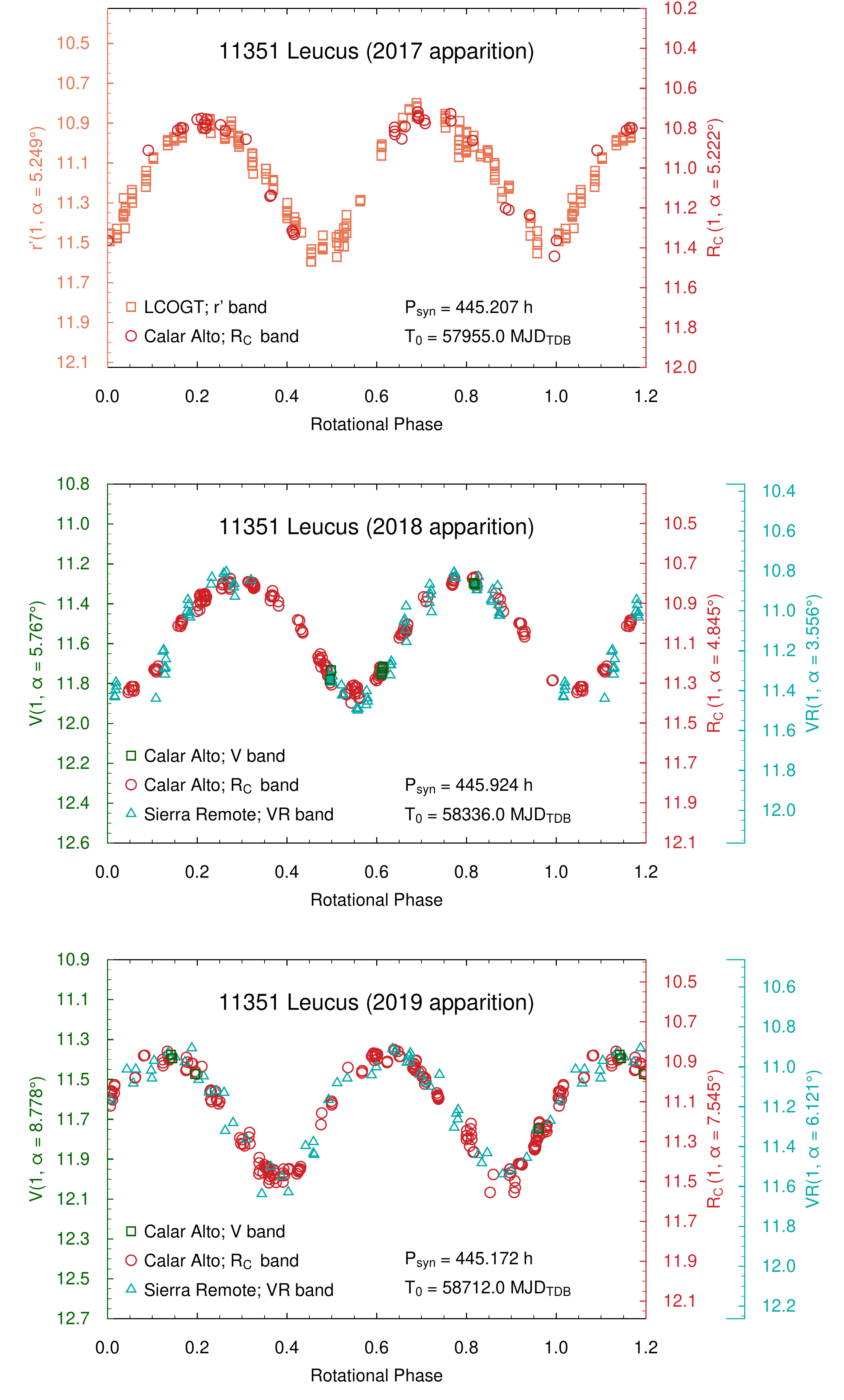}
\caption{Composite lightcurves for the 2017, 2018 and 2019 apparitions. The data points are folded with the synodic periods
listed in the respective graphs, with zero phase corresponding to the respective T$_0$ epochs. T$_0$ are one-leg, light-travel time corrected 
Modified Julian Dates (MJD) expressed in the TDB uniform time frame. The listed synodic periods
are the exact numbers used for folding the composites, and as such, are reported without uncertainty.
Data points beyond rotation phase 1.0 are repeated for clarity. 
The magnitudes are reduced to 1 au from the observer and from the Sun and to the respective reference phase angles
by using a nominal linear phase coefficient $\beta=0.0395$ mag/$\degr$.\label{fig:compo-fig}}
\end{figure}

The composite lightcurves for the 2017, 2018 and 2019 apparitions are reported in Figure \ref{fig:compo-fig}.
It can be seen that the respective synodic periods differ among each other by as much as 0.7 h, which corresponds to about 0.15\%.
This is expected for two lines of reasons. First, due to the slow rotation, the useful baseline for the determination of the periods for each apparition just covers   
a handful of rotational cycles, which limits the accuracy of the determination. Secondly -- and to a lesser extent -- the apparent instantaneous rotation rate depends on 
the rate of change of the topocentric ecliptic longitude of the object, which can make the synodic period change slightly 
\replaced{from apparition to apparition}{during the course of an appartition and
from one apparition to the next}. 
In the next sections we will derive
a very accurate sidereal period and phase function, by using a dynamical and shape model that makes use of all of the available observations,
over a baseline of about 6 years.

\begin{deluxetable*}{ccccccccccc}
\tablenum{1}
\tablecaption{Observational circumstances\label{tab:obscirc}}
\tablewidth{0pt}
\tablecolumns{11}
\tablehead{
\colhead{Date} & \colhead{$\lambda$} & \colhead{$\beta$} & \colhead{$\alpha$} &\colhead{r} & \colhead{$\Delta$} &  \multicolumn{2}{c}{$\lambda$ (PAB) $\beta$}  & \colhead{Band}  & \colhead{Observatory}  & \colhead{Observers}\\
\colhead{(UT)} & \multicolumn{2}{c}{($\degr$  J2000)} & \colhead{($\degr$)} &\colhead{(au)} & \colhead{(au)} & \multicolumn{2}{c}{($\degr$  J2000)}
}
\startdata
2017 May  1.1 & 287.4 & +5.7 & 9.683 & 5.5071 & 5.0320 & 282.5 & +5.5 & R$_C$& 493& SH, SMo \\
2017 May  2.1 & 287.4 & +5.8 & 9.616 & 5.5067 & 5.0175 & 282.6 & +5.5 & R$_C$& 493& SH, SMo \\
2017 May 20.1 & 287.0 & +6.3 & 7.887 & 5.5005 & 4.7745 & 283.0 & +5.9 & R$_C$& 493& SH, SMo \\
2017 May 28.1 & 286.5 & +6.6 & 6.841 & 5.4977 & 4.6852 & 283.1 & +6.1 & R$_C$& 493& SMo, SH \\
2017 May 30.1 & 286.3 & +6.6 & 6.550 & 5.4970 & 4.6646 & 283.1 & +6.1 & R$_C$& 493& SMo, SH \\
2017 May 31.0 & 286.3 & +6.6 & 6.408 & 5.4966 & 4.6551 & 283.1 & +6.1 & R$_C$& 493& SMo, SH \\
\hline
2017 Jul 16.4 & 281.0 & +7.6 &  2.738 & 5.4798 & 4.4914 & 282.2 & +6.9 & r' & Q64	& MWB, AZ \\
2017 Jul 17.0 & 280.9 & +7.6 &  2.833 & 5.4795 & 4.4932 & 282.1 & +6.9 & r' & W86	& MWB, AZ \\
2017 Jul 17.3 & 280.9 & +7.6 &  2.882 & 5.4794 & 4.4942 & 282.1 & +6.9 & r' & W85	& MWB, AZ \\
2017 Jul 17.6 & 280.8 & +7.6 &  2.933 & 5.4793 & 4.4952 & 282.1 & +6.9 & r' & Q63	& MWB, AZ \\
2017 Jul 17.9 & 280.8 & +7.6 &  2.980 & 5.4792 & 4.4962 & 282.1 & +6.9 & r' & K93	& MWB, AZ \\
2017 Jul 18.2 & 280.8 & +7.6 &  3.031 & 5.4791 & 4.4973 & 282.1 & +6.9 & r' & W86	& MWB, AZ \\
\hline
2018 Jun 11.1 & 318.3 & +11.1 & 9.382 & 5.3394 & 4.7451 & 313.5 & +10.5 & R$_C$ & 493 & SH, SMo\\
2018 Jul  9.1 & 316.6 & +12.1 & 5.829 & 5.3263 & 4.4379 & 313.8 & +11.1 & V & 493 & SH, SMo\\
2018 Jul 15.0 & 316.0 & +12.2 & 4.895 & 5.3236 & 4.3947 & 313.8 & +11.2 & V & 493 & SH, SMo\\
2018 Sep  4.9 & 309.8 & +12.5 & 6.171 & 5.2990 & 4.4362 & 312.8 & +11.5 & V & 493 & SH, SMo\\
2018 Aug  6.9 & 313.2 & +12.7 & 2.406 & 5.3127 & 4.3186 & 313.3 & +11.5 & R$_C$ & 493 & SH, SMo\\
2018 Aug  8.0 & 313.1 & +12.7 & 2.432 & 5.3123 & 4.3187 & 313.3 & +11.5 & R$_C$ & 493 & SH, SMo\\
2018 Aug  9.0 & 312.9 & +12.7 & 2.475 & 5.3118 & 4.3192 & 313.3 & +11.5 & R$_C$ & 493 & SH, SMo\\
\hline
2019 Nov  9.1 & 342.2 & +12.6 & 10.103 & 5.1016 & 4.5976 & 347.3 & +12.0 & VR & G80 & MWB\\
2019 Nov 10.1 & 342.2 & +12.6 & 10.181 & 5.1012 & 4.6112 & 347.4 & +12.0 & VR & G80 & MWB\\
2019 Nov 11.1 & 342.2 & +12.5 & 10.257 & 5.1008 & 4.6249 & 347.4 & +12.0 & VR & G80 & MWB\\
2019 Nov 12.1 & 342.2 & +12.5 & 10.329 & 5.1004 & 4.6387 & 347.5 & +11.9 & VR & G80 & MWB\\
2019 Nov 24.1 & 342.6 & +12.0 & 10.962 & 5.0955 & 4.8123 & 348.2 & +11.7 & VR & G80 & MWB\\
\enddata
\tablecomments{This table is an excerpt. The observational circumstances for all of the observation nights are reported in the online material.
$\lambda$ and $\beta$ are the topocentric ecliptic longitude and latitude of the target, respectively.
$\alpha$ is the solar phase angle, r is the heliocentric distance and $\Delta$ is the topocentric range of Leucus.
$\lambda$ and $\beta$ (PAB) are the topocentric ecliptic longitude and latitude of the phase angle bisector, as defined in \citet{Harrisetal1984} 
} 
 \end{deluxetable*}


\section{Modeling} \label{sec:modeling}
\subsection{Data} \label{subsec:data}

The photometric lightcurves presented in the previous section and in \citet{Buieetal2018},
and the results from the stellar occultation campaigns reported in \citet{Buieetal2020}
constitute the bulk of observational data used for our modeling work. 
In addition, we made use of dense lightcurve photometry by \citet{Frenchetal2013}
(retrieved through the ALCDEF service (\url{http://alcdef.org/})) and sparse photometry from the following sources:

\begin{enumerate}
\item The Gaia DR2 database \citep{GaiaCollaboration2018} retrieved through the VizieR server (\url{https://vizier.u-strasbg.fr/}),
\item The ZTF project \citep{Bellmetal2019} retrieved through the IRSA server (\url{https://irsa.ipac.caltech.edu/applications/ztf/})
and from the nightly transient archive at \url{https://ztf.uw.edu},
\item  The PAN-STARRS-1 DR2 database \citep{flewellingetal2016} retrieved through a query to the  MAST archive (\url{https://catalogs.mast.stsci.edu/}), and
\item The ATLAS project \citep{Tonryetal2018} retrieved through the AstDys database (\url{https://newton.spacedys.com/astdys/}).
\end{enumerate}

\added{For all but the ATLAS observations, the photometric uncertainties were retrieved along with the magnitude data.
In the case of the ATLAS observations, we estimated nominal photometric uncertainties by compiling the data into 
composites for the individual oppositions and computing their residuals.  
}
Since the survey observations were acquired in a variety of different photometric systems, for which transformations to the 
Johnson system are not accurately established, they were treated as relative photometry. 
\added{Similarly to the dense data set, the times were light-time corrected, and the magnitudes reduced to 1 au.}
Although these additional datasets provide varying degrees of accuracy, they proved to be very useful
to extend the coverage and baseline of the observations, which combined, cover a period of about 6 years.\\

\subsection{Convex inversion} \label{subsec:convexinversion}

In this paper we apply the convex shape inversion approach described in \citet{Kaasetal2002book} and references therein to the photometric time series
to simultaneously solve for the sidereal period, the spin axis orientation, the photometric function and a convex, polyhedral approximation of the shape. 
The occultation data are used as a constraint to resolve the spin axis ambiguity, to determine the scale of the object -- and hence its albedo -- and to refine the orientation of the spin axis.
Although it is likely that Leucus does contain some degree of global-scale concavity, we did not feel that
the available data -- both because of coverage and photometric accuracy, in the case of lightcurve data, and because of
limited unique observation geometries in the case of occultation data -- would allow addressing of the intrinsic 
non-uniqueness of the non-convex problem. On the other hand, the convex inversion scheme offers the advantage of a provably convergent method that 
results in a unique solution in the case of a convex shape \citep{KaasalainenandLamberg2006}, and gracefully degrades in the case of moderate concavities.
Non-convex modeling of Leucus will be the subject of future work as soon as more data -- especially 
at further occultation geometries -- become available. 
\added{Although a working implementation of the convex inversion algorithm (\textit{convexinv}) is publicly available \citep{Durechetal2010}
we decided to develop our own implementation based on the original publications \citep{KaasalainenandTorppa2001,Kaasalainenetal2001}.
This approach has allowed us to overcome some limitations of the code (described later), to expand its functionality, and to correct a minor
 bug in the computation of the $\chi^2$ metric that is present in the current \textit{convexinv} version and that has been promptly
 communicated to the author. Although we did not make use of any code from \textit{convexinv}, we did use that software
to validate the results of our own code on a few test cases.
}\\

The surface brightness of the object is described through its photometric function, which, for the purpose 
of this work, is assumed to be separable into a \textit{disk function} and a \textit{surface phase function} (see e.g. \citealt{Schroederetal2013}). 
Following \citet{Kaasalainenetal2001}, we adopt Lommel-Seelinger-Lambert (LSL) scattering as a disk function and 
a 3-parameter exponential-linear combination as a surface phase function, as described in Appendix A. The LSL disk function is the linear combination 
of the Lommel-Seeliger and Lambert scattering functions through a partition parameter $c$, and is equivalent to the Lunar-Lambert
disk function (\citealt{Lietal2015} and references therein) when the latter is used with a partition function independent on the phase angle. 
We fixed $c$ parameter to a constant value $c=0.1$, appropriate for dark asteroids \citep{Kaasalainenetal2005}.
For this work we didn't attempt to use more complex photometric models -- as e.g. the Hapke function (\citealt{Hapke2012} and references therein) --  because, due to the small phase angle range in which
Trojans can be observed from Earth, no meaningful retrieval of the model parameters is achievable.\\

The brightness of the object depends on the product of its size and its albedo. From unresolved photometric measurements alone,
it is not possible to retrieve independently those two quantities. Independent measurements as thermal radiometry, stellar occultations or direct imaging,
however, offer the possibility to disentangle the two quantities. 
In absence of more detailed information, we assume \deleted{that} that the photometric properties of the target -- in particular its albedo --
are uniform over its surface. This is quite a reasonable assumption to be made, because, although albedo variations on asteroids do occur, 
the observed lightcurve variations tend to be dominated by the changing cross section of a non-spherical shape.\\

The convex inversion scheme models the brightness of the target in the space of its Extended Gaussian Image (EGI, \citealt{KaasalainenandTorppa2001}), 
as opposed to the 3D object space. The EGI represents the discrete equivalent of the inverse of the Gaussian Surface Density, and is used to represent 
the area of the facets oriented towards a particular direction. For the Leucus model we 
parametrize the EGI with a spherical harmonic expansion of rank and order 
\replaced{6, which results in a total of 49}{8, which results in a total of 81} shape parameters, one of which represents its scale. An exponential-function representation of the EGI is used for 
guaranteeing the positiveness of the surface areas.  
We integrate the EGI over the unit sphere by sampling the spherical harmonics expansion at discrete points by following a 
Lebedev quadrature \citep{LebedevandLaikov1999}, which offers greater efficiency compared to a quadrature based, e.g. 
on a uniform or random sampling \citep{Kaasalainenetal2012}. In the case of Leucus we applied a Lebedev rule of order 302, which
results in an EGI with the same number of facets.\\

\added{Contrary to \citet{KaasalainenandTorppa2001} \added{-- and to the \textit{convexinv} implementation --} we minimize a weighted metric for solving the least-squares problem,
in order to properly account for the varying degree of accuracy of the different data sets.}
In the case of absolute observations, the optimization is performed by minimizing the reduced $\chi^2_{red}$ metric defined as 

 \begin{equation} 
 \chi_{red}^{2}= \frac{1}{N}\sum\limits_{i}{\frac{(L_i^{obs}-L_i^{mod})^2}{\sigma_i^2}}
   \end{equation} 

where  $L_i^{obs}$ and $L_i^{mod}$ are the observed and model intensities, respectively, $\sigma_i$ are the intensity uncertainties
and $N$ is the number of degrees of freedom for errors. The index $i$ runs over all of the $n$ photometric data points.\\
In the case of relative observations, we minimize the deviations of the intensities relative to the average of the respective lightcurve:

 \begin{equation} 
 \chi_{rel}^{2}= \frac{1}{N}\sum\limits_{k,j}\left(\frac{\bar{L}^{obs}_{j}}{\sigma_{k,j}}\right)^2\left(\frac{L_{k,j}^{obs}}{\bar{L}^{obs}_{j}}-\frac{L_{k,j}^{mod}}{\bar{L}^{mod}_{j}}\right)^2
   \end{equation} 

where the index $k$ runs over the data points of the lightcurve $j$ and $\bar{L}^{obs}_{j}$ and $\bar{L}^{mod}_{j}$ are the average intensities of the
observed and modeled lightcurve $j$, respectively. \added{The term $\sigma_{k,j}$ represents the uncertainty of the $k^{th}$ data point of lightcurve $j$.}
If absolute observations were performed during the same apparition in two photometric bands, then
a color index term is introduced to tie the two photometric systems. The color term is also optimized in the procedure.
If, on the other hand, one photometric band is never used together with another photometric system
at least during one apparition, then these series of observations are treated as relative photometry, in order to avoid a possible parameter coupling 
between the color index and the spin axis orientation. The non-linear optimization is performed with the Levenberg-Marquardt algorithm \citep{NumericalRecipes}.\\

A unique transformation from the EGI to a polyhedron in 3D space is guaranteed by a Minkowski theorem \citep{Minkowski1897}.
For this transformation we use the iterative scheme proposed by \citet{Lamberg1993}.
The body reference system is defined such that the Z axis coincides with the spin axis. The plane that contains the body center of mass
and that is perpendicular to the Z axis defines the XY plane in the body system. The direction of the X axis is chosen such that it coincides with the projection of the
principal axis of smallest inertia onto the XY plane. The body X axis also defines the location of the prime meridian and thus the zero longitude.
 

\subsection{Rotation model} \label{subsec:rotationandshape}

\begin{figure}
\plotone{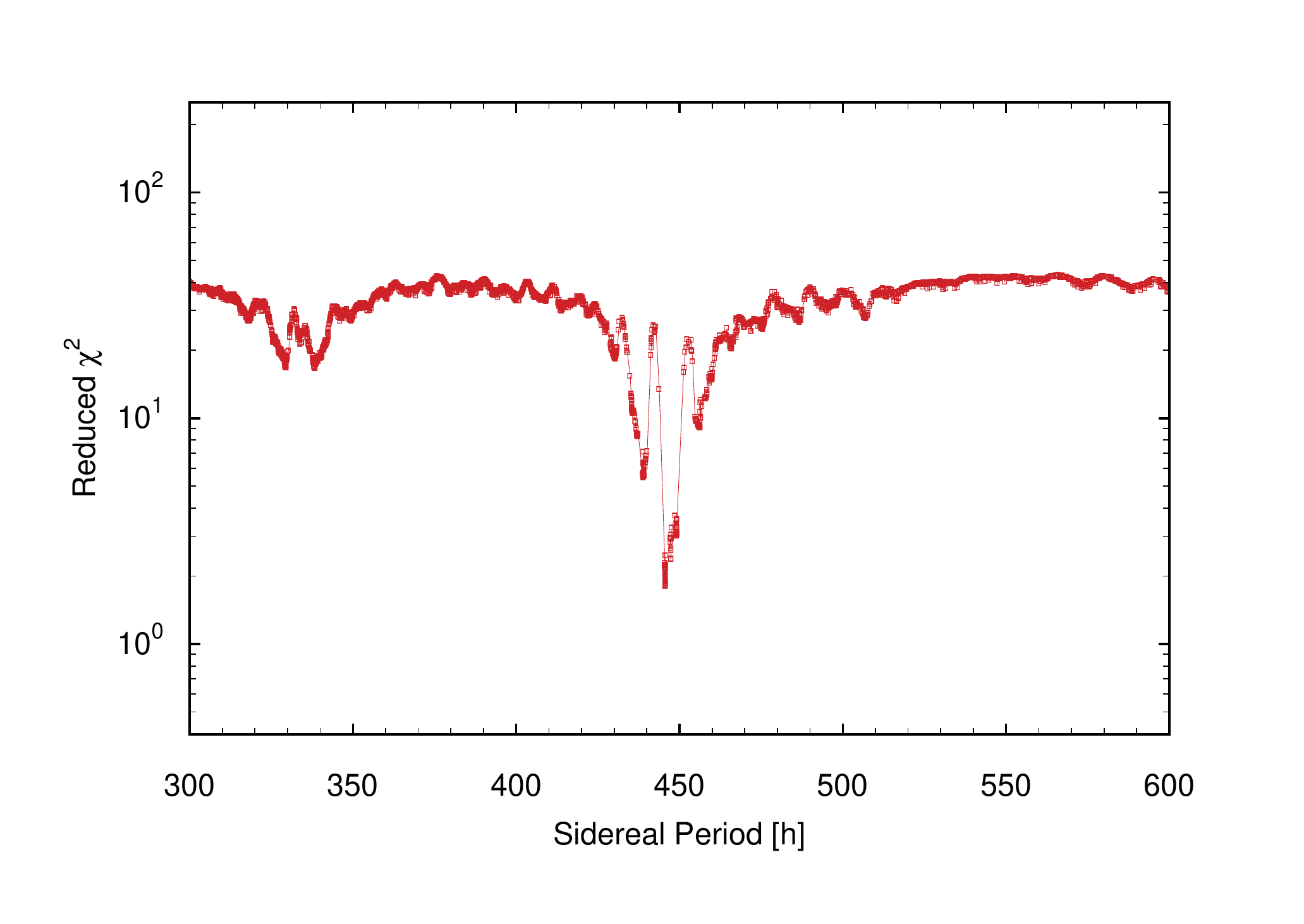}
\caption{Result of the sidereal period scan. 
\added{Each data point corresponds to a local-minimum solution
obtained by using trial periods in the
range between 300 h and 600 h as iteration start values for the sidereal period.
The global minimum around 445.7 h corresponds to the best-fit solution.}
\label{fig:period-fig}}
\end{figure}

The rotation period strongly modulates the spectrum of the $\chi^2$ of the fit with periodic local minima at 
a minimum spacing $\Delta P\approx P^2/(2T)$ \citep{Kaasalainenetal2001}, where $P$ is the rotation period and $T$ is the total 
baseline of the lightcurve coverage.
It is therefore important that the optimization be started in the vicinity of the correct 
period, in order to avoid that the optimizer could become trapped in the local minimum of an alias period.
For this reason, the search of the correct sidereal period is the first step in the convex inversion scheme.
The search is performed by running the inversion procedure by using as starting conditions all trial periods 
in a relevant range, with a step sufficiently smaller than the minima separation.
For each trial period we use twelve different starting pole directions.
Figure \ref{fig:period-fig} shows the results of the period scan for Leucus.  

\begin{figure}
\plotone{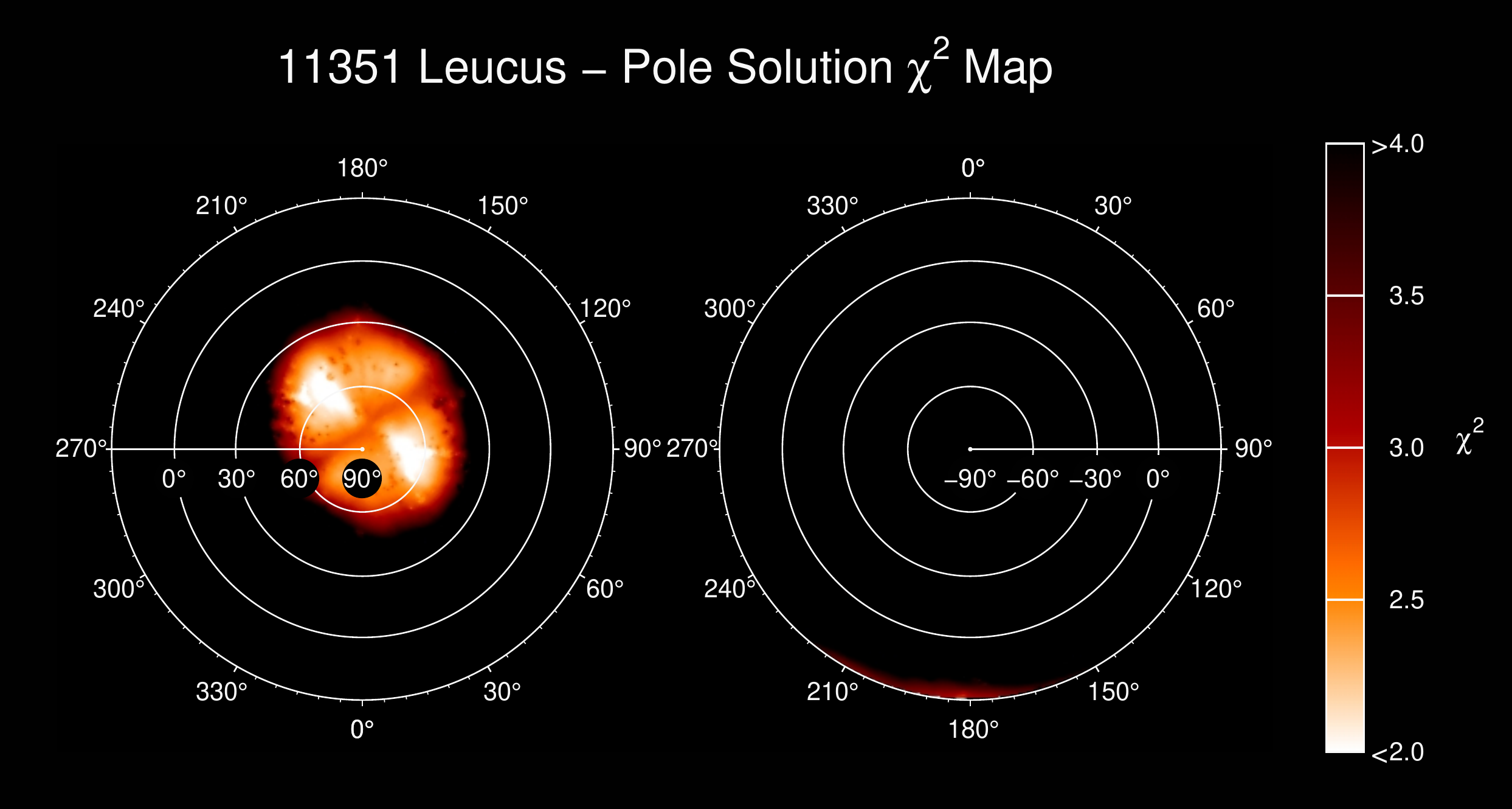}
\caption{Polar azimuthal equidistant projection of the $\chi^2$ of the pole solutions. 
The coordinates are expressed in the J2000 Ecliptic Frame. 
The left panel is centered on the North Ecliptic pole and the right one on the South Ecliptic pole. 
The loci of the two complementary best solutions are clearly visible as white regions. \label{figchi2map-fig}}
\end{figure}

In order to identify the coarse direction of the spin axis we run the optimization procedure 
by using the best-fit sidereal period derived in the previous \replaced{section}{paragraph} as a starting value and by
fixing the spin axis orientation to each of about 20,000 trial directions 
equally spaced on the celestial sphere. The shape of the object, the sidereal period 
and the photometric parameters (but not the pole coordinates) are simultaneously
optimized for each trial pole. The resulting $\chi^2$ values for each solution are mapped 
on the celestial sphere via a polar azimuthal equidistant projection and are shown in Figure \ref{figchi2map-fig}.
As expected, two, equally significant loci for the best solution are identified. This is the consequence of the
\textit{ambiguity theorem} \citep{KaasalainenandLamberg2006} that states that if disk-integrated 
photometric observations are always carried out in the same photometric plane -- as is the case for 
low-inclination objects observed from Earth -- then 
two indistinguishable solutions exist that satisfy the observations and that are separated by 
about 180$\degr$ in Ecliptic longitude.
The shapes corresponding to the two solutions are approximately mirrored shapes of one 
another around the body XY-plane.
The graph also shows that both solutions are prograde and, as already inferred by 
\citet{Buieetal2018}, the obliquity of the spin axis is low.\\

\subsection{Fit to the occultation data} \label{subsec:occultationfit}

By providing disk-resolved information, occultation data can resolve the pole ambiguity,  
fix the absolute scale of the shape model and, together with the determined $H_V$-value, 
measure its geometric albedo.\\
In principle, occultation data can also be used to derive non-convex shape models,
provided sufficient, densely sampled silhouettes are available at multiple rotation phases.
Much work has been recently done concerning the optimal fusion of data coming
from disk-integrated photometry and disk-resolved techniques as stellar occultations, adaptive-optics
direct imaging and interferometry (e.g. \citealt{KaasalainenandViikinkoski2012,Viikinkoskietal2015}). 
In the case of Leucus, however, both the limited lightcurve coverage
and the sparse silhouette sampling would not allow a reliable non-convex model to be derived.
For this reason we decided to adopt an approach similar to \citet{Durechetal2011} and favor the 
advantages of the uniqueness and stability of a convex solution. \\

For each of the two best candidate solutions from the 
previous section
we project the vertices of the shape model onto the 
plane of sky at the time of each occultation event and then compute the 2D convex
hull of the projected points. 
Applying this procedure to the two complementary solutions visible in Fig. \ref{figchi2map-fig} allowed us to unambiguously 
identify the correct solution as the one centered at an Ecliptic longitude of around 210$\degr$. However,
it also became apparent that the best solution had a slight systematic deviation in the orientation
of the projection with respect to the occultation data that could be explained by a slight offset ($\approx$5$\degr$)
in the direction of the spin axis orientation of the model. Such a small mismatch was not unexpected, as the loci of the solutions
in Fig. \ref{figchi2map-fig} are quite broad and shallow, and a small shift in the spin axis direction of the model would produce
fits to the lightcurves with similar $\chi^{2}$. On the other hand, disk-resolved data as stellar occultations are much more sensitive
to a pole misalignment. For this reason we decided to use the occultation data for the refinement of the solution.

As a goodness of fit for the occultation data we 
define a metric

 \begin{equation} 
\chi_{occ}^{2}= \sum\limits_{i,j}\frac{{(D_{ij})^2}}{N_{op}}
   \end{equation} 

where $D_{ij}$ represents the minimum Euclidean distance between the occultation transition point $j$ (either ingress or egress) of the event $i$
and the model 2D convex hull of the event $i$.  $N_{op}$ is the total number of the observed transition points. This metric is different from the one
chosen by \citet{Durechetal2011}, who prefer to use the distance of the occultation points to the
occultation limb measured in the direction of the asteroid ground track. Their choice is justified by the fact that 
the largest contribution to their occultation data is given by timing errors and observer reaction times,
which act along track. In our case, on the other hand, we estimate the largest errors to be due to  
the convex shape model, which are not expected to have a preferred direction. \\

  $\chi_{occ}^{2}$ is minimized by optimizing the global scale and the Cartesian coordinates of the 
  centers of the projections for each occultation epoch.
The latter is necessary to compensate both for inaccuracies in the position of the occulted stars and for 
the uncertainties of the target ephemeris at the epochs of the 
occultations. In practice the minimization is performed
by varying the projection centers and the $A_{LSL}$ albedo (see Appendix A) -- which 
constrains the scale -- in an adaptive grid search fashion.\\

One practical problem arises from the fact that the convex shape optimization is performed in the EGI space, while
the occultation profile optimization is done in the space of the projected shape. It is therefore impractical to 
perform a simultaneous, combined optimization. For this reason we used the $\chi_{occ}^{2}$ as a mild
penalty function for a combined $\chi_{tot}^{2}$ of the form

 \begin{equation} 
 \chi_{tot}^{2}= \chi_{conv}^{2} + \lambda \chi_{occ}^{2}
   \end{equation} 

where $\chi_{conv}^{2}$ is the term coming from the convex inversion and $\lambda$ 
is a small weight. The value of $\lambda$ has been determined by experimentation
by adopting a value of $\lambda$ that minimized $\chi_{occ}^{2}$
without significantly increasing $\chi_{conv}^{2}$.
We have then computed the quantity $\chi_{tot}^{2}$ for hundred discrete pole directions 
within a radius of 10$\degr$ of the best solution (and of the complementary one).\\

\begin{figure}
\plotone{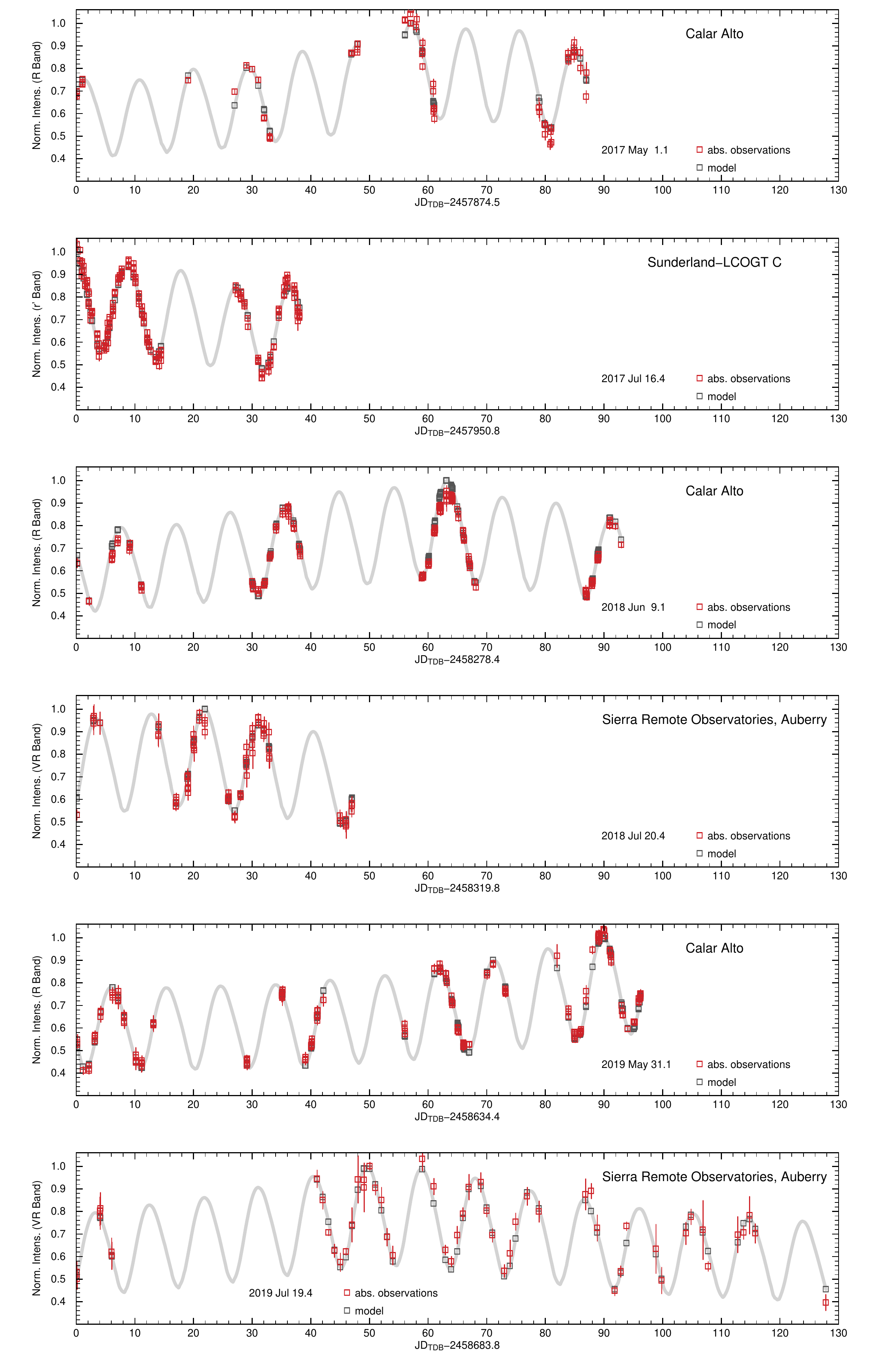}
\caption{Observed lightcurves shown along with the respective model lightcurves. 
The data are plotted in intensity and 
are normalized to unity at the model maximum value for the respective opposition. The dates 
reported in the labels represent the epoch of the first observation of each sequence.\label{fig:modlcs}}
\end{figure}

The parameters for the best-fit solution are reported in Table \ref{tab:results}.
The errors quoted for the different quantities have been determined by computing 
perturbed solutions and investigating the effects on the $\chi^{2}$. It must be noted that
this method produces an evaluation of the statistical component of the uncertainties only. 
The largest contribution \added{to} the errors is thought to be due to the violations of the 
assumptions, as the assumption of the functional form of the photometric model 
and the convexity assumption. Those contributions, however, are virtually impossible to be formally
quantified. For this reason we refrained from using more sophisticated statistical error models,
as the Markov chain Monte Carlo (MCMC) method, because they also only address the stochastic 
component of the uncertainty.

Figure \ref{fig:modlcs} shows the synthetic lightcurves for the corresponding shape model for the epochs covered by 
the dense observations reported in this paper.  
The lightcurve intensities are corrected for changing heliocentric and topocentric range and are normalized to unity at the maximum 
of the respective observation window. Besides the rotational variation, the intensity variation due to the changing phase angle is visible.

\begin{figure}
\plotone{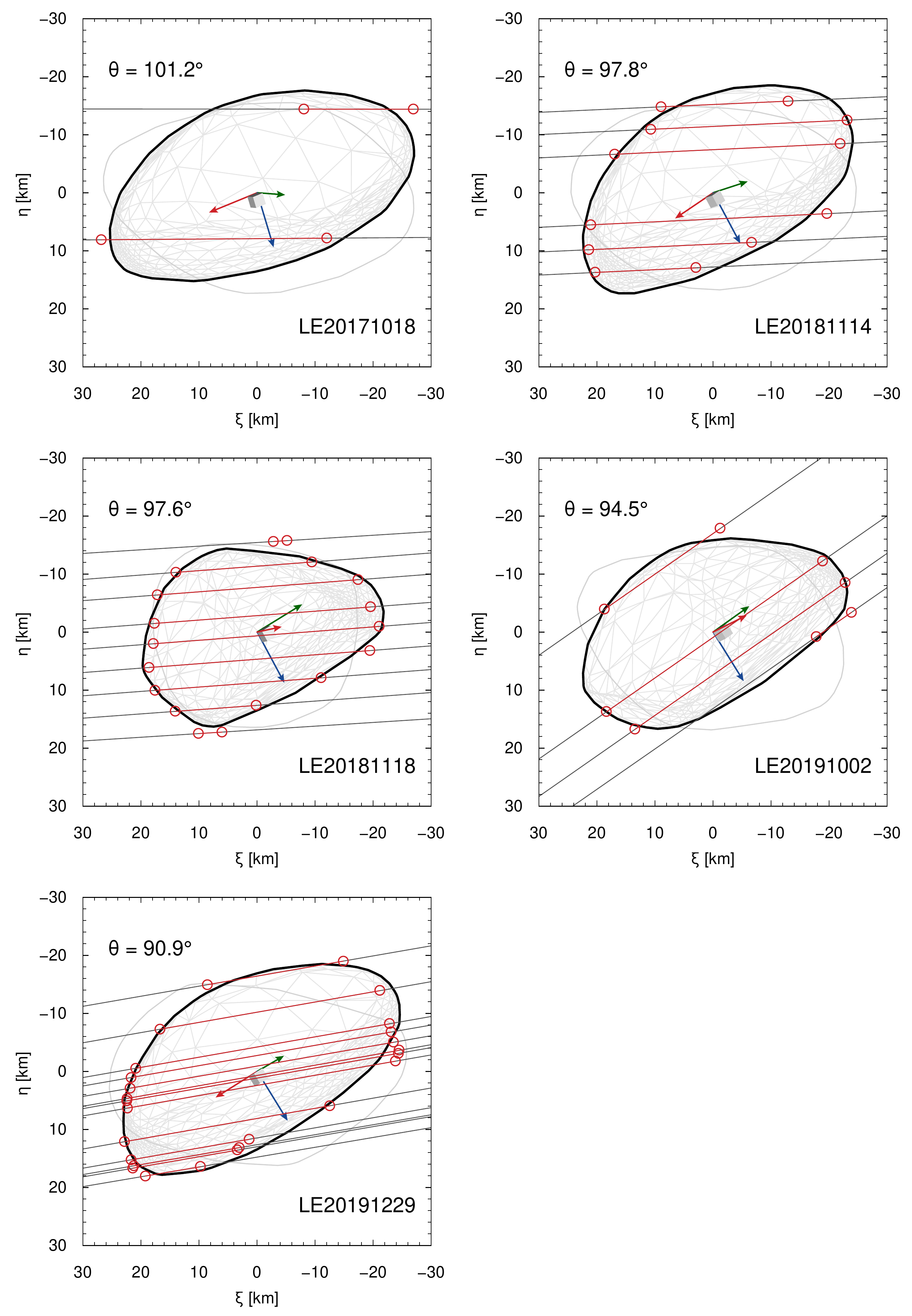}
\caption{Occulting silhouettes of the best-fit convex model of Leucus (solid black line) along 
with the rejected complementary model displayed in light gray. 
The five panels correspond to the epochs of the five occultation events reported in \citep{Buieetal2020}.
The red circles correspond to the starts and ends of the respective positive occultation chords.  
The red, green, and blue arrows represent the X, Y, and Z axes in the body-fixed reference frame, respectively.
$\theta$ represents the aspect angle, i.e. the angle between the spin axis and the target-observer direction.
\label{fig:silhouette-fig}}
\end{figure}

Figure \ref{fig:silhouette-fig} shows the resulting best-fit model overplot onto the occultation chords by \citet{Buieetal2020}.
The complementary (wrong) solution is also plotted in light gray, showing how occultation data can help identify
the correct solution. Please note that, for comparison purposes, we plot the data following Buie's convention of
projecting the shape onto the plane of sky \citep{Green}, with the $\eta$ coordinate increasing towards celestial North
and the $\xi$ coordinate increasing towards East. This is a different convention than in \cite{Durechetal2011},
who project the shape onto the fundamental plane, with the $\eta$ coordinate increasing towards 
West and the $\xi$ coordinate increasing towards North.

We note that the model shape well reproduces the occultation profiles and
captures the general shape of the object, within the limits of a convex representation. In particular,
it reproduces well the flat sides visible during the events LE20181118 and LE20191002 and the 
polygonal appearance of event LE20181118.
It is also important to note that the occultation data are not directly used to derive the convex shape. The shape is
influenced by the occultation data only indirectly, through the refinement of the spin axis 
orientation. Under this light, the match of the convex shape model to the occultation data appears even more convincing.

\begin{figure}
\plotone{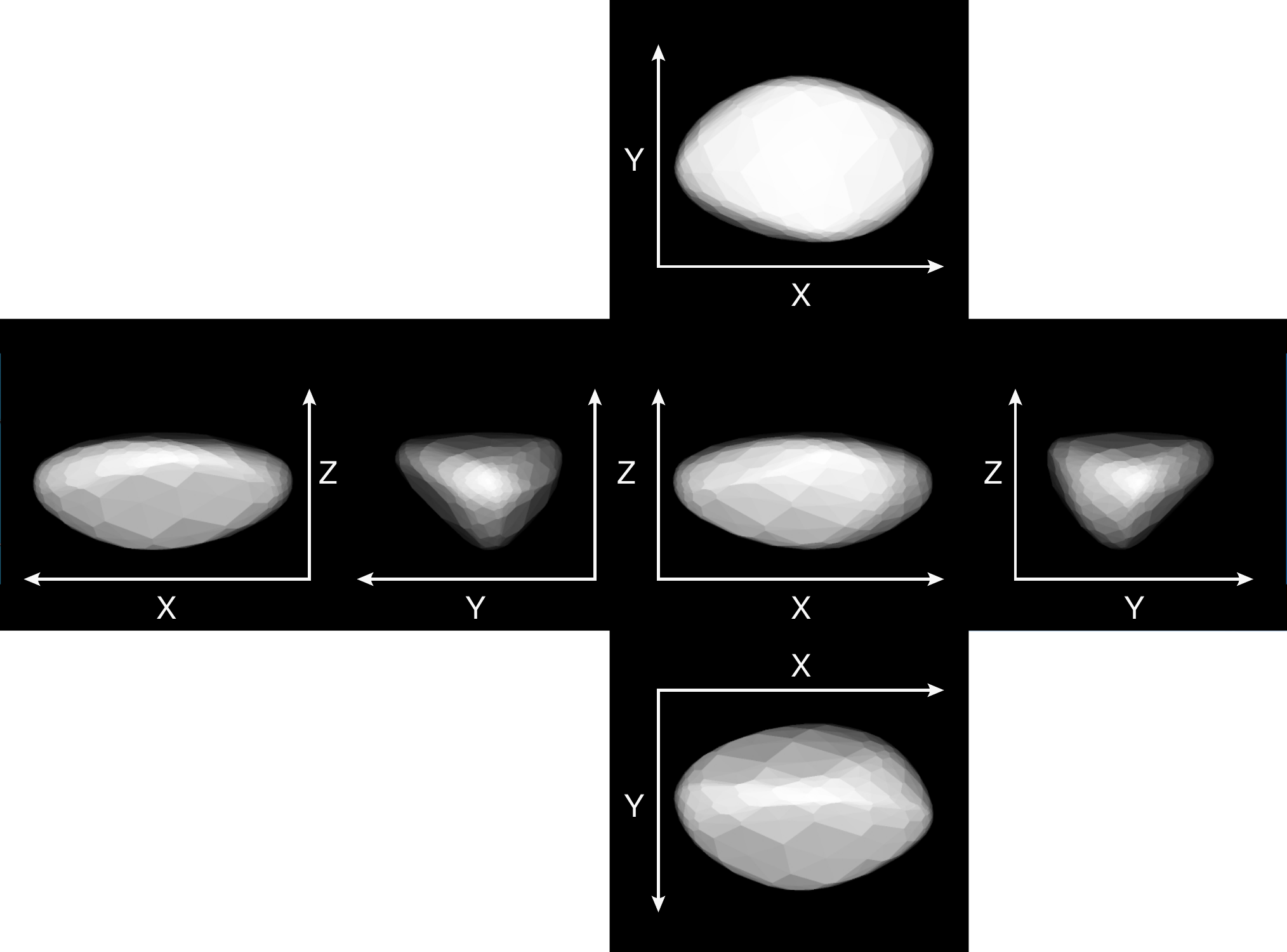}
\caption{Six orthogonal projections of the best-fit Leucus convex shape arranged similarly to an unfolded dice.
For contrast reasons, Lambert shading is used for the figure rendering instead of the LSL disk function used in the modeling.  \label{fig:dice-fig}}
\end{figure}

Figure \ref{fig:dice-fig} shows six orthogonal views of the best-fit Leucus shape model. As already hinted by the occultation silhouettes, 
Leucus' shape considerably deviates from an ellipsoid and is characterized by a comparatively flat Northern hemisphere.  \deleted{and an
approximately triangular cross section in the body YZ plane.} 
\added{The convexity residual of the shape model resulting from the Minkowski 
transformation \citep{KaasalainenandTorppa2001} is a mere 0.14\%, which, taken at face value, would suggest negligible
global-scale concavities. However, the small phase angle range at which observations are available reduces the diagnostic value of this parameter.}

As a sanity check, we computed the model's inertia tensor with the method by \citet{Dobrovolskis1996} -- assuming 
a uniform bulk density -- and established that the principal inertia axis is misaligned with 
respect to the body's rotation axis by about 8$\degr$. To this end, we have to recall that the 
derived shape model represents a convex approximation of the shape,
and ignoring concavities can contribute to shift the direction of the inertia axes. Also the 
assumption of uniform bulk density, if violated, would
contribute to shift the direction of the principal axis of inertia. The observed misalignment is therefore not necessarily a hint that the object 
is not in a principal rotation state. Rather, it is an expression of the fact that the convex shape represents a \textit{photometric shape}, which
can locally differ from the physical shape. 

The maximum extent of a complex shape can be defined in several different ways (see e.g. \citealt{Torppaetal2008}).
For our model we define the maximum dimensions as

\begin{equation} 
\begin{array}{ccc}       L_X=max(x_1, ..., x_i)-min(x_1, ..., x_i)  \\L_Y=max(y_1, ..., y_i)-min(y_1, ..., y_i) \\L_Z=max(z_1, ..., z_i)-min(z_1, ..., z_i)  \end{array} 
\end{equation} 
where $(x_i,y_i,z_i)$ represent the Cartesian coordinates of the $i^{th}$ vertex of the shape model 
and the $X$, $Y$, and $Z$ body axes are defined as per Sec. \ref{subsec:convexinversion}.
This definition produces similar -- but not identical -- extents as the \textit{Overall Dimensions} (OD) definition of \citet{Torppaetal2008} (see their Fig. 1).
In particular, with our definition the largest extent is computed in the general direction of the principal axis of smallest inertia, which represents 
a natural axis of the body. In the case of the OD definition by \citet{Torppaetal2008}, on the other 
hand, the maximum dimension is the largest extent that occurs anywhere in the XY plane. 
As an example, for a hypothetical body with a rectangular equatorial cross section, with our definition 
the maximum extent would be represented by the longest side of the rectangle, while, according to the 
definition by Torppa the maximum extent would be represented by the diagonal of the rectangle.

With our definition, the maximum dimensions for Leucus are $L_X=60.8$ km,  $L_Y=39.2$ km and  $L_Z=27.8$ km (see Table \ref{tab:results}).
Those compare with the axes  (63.8, 36.6, 29.6) km of the ellipsoidal approximation by \citet{Buieetal2020}
that were derived under the assumption of a strictly equatorial aspect during the occultation events.
In the same table the orientation of the model is described both by reporting the Ecliptic J2000 coordinates of the spin axis and 
initial angle $\Phi_0$ at the epoch $T_0$ using the formalism by 
\citet{Kaasalainenetal2001}, as well as using the IAU convention of reporting the ICRF equatorial coordinates of the spin axis 
and the $W_0$ angle at the standard epoch J2000 (\citealt{Archinaletal2018,Archinaletal2019}).

The surface-equivalent spherical diameter of the convex model is $D=41.0\pm0.7$ km, whereas its volume is 
$(30\pm2)\times10^3 ~\rm km^3$. It should be noted, however, that due to the likely presence of concavities 
the quoted value for the volume rather represents an upper bound.


\begin{deluxetable*}{cc}
\tablenum{2}
\tablecaption{Results\label{tab:results}}
\tablewidth{0pt}
\tablecolumns{2}
\tablehead{
}

\startdata
Sidereal Period (h) &  445.683 $\pm$ 0.007\\
Pole J2000 Ecl. Longitude ($\degr$) & 208 \\
Pole J2000 Ecl. Latitude ($\degr$) & +77  \\
Pole J2000 RA  ($\degr$) & 248  \\
Pole J2000 Dec  ($\degr$) & +58\\
Radius of pole uncertainty ($\degr$, 1$\sigma$) &3\\
Ecliptic obliquity of pole ($\degr$) & 13\\
Orbital obliquity of pole ($\degr$) & 10  \\
$T_0$ (JD$\rm _{TDB}$) & 2456378.0 \\
$\Phi_0$  ($\rm \degr$)   & -76.129 \\
$W_0$ ($\degr$)  & 60.014 \\
$\dot{W}$ ($\rm\degr day^{-1}$)  & 19.38596 $\pm$ 0.00030\\
$p_{\rm V}$ & 0.043  $\pm$ 0.002\\
$A_{\rm LSL}$ & 0.061  $\pm$ 0.002\\
$A_0$ & 0.23  $\pm$ 0.09\\
$D$ (rad)& 0.075  $\pm$ 0.015\\
$k$ ($\rm rad^{-1}$)& -1.07  $\pm$ 0.23\\
$c$ (fixed)& 0.1  \\
$H_{\rm V-LSL}$ (sph. int.)& 10.979  $\pm$ 0.037\\
$H_{\rm V-lin}$ (sph. int.)& 11.034  $\pm$ 0.035\\
$\beta$ ($\rm mag/\degr$) & 0.0395 $\pm$ 0.005\\
$H_{\rm V-HG}$ (sph. int.)& 10.894  $\pm$ 0.004\\
$G$ & 0.34  $\pm$ 0.02\\
$H_{\rm V-HG_1G_2}$ (sph. int.)& 10.95  $\pm$ 0.01\\
$G_1$ & 0.63  $\pm$ 0.04\\
$G_2$ & 0.23  $\pm$ 0.02\\
$V-R$& 0.464  $\pm$ 0.015\\
$V-r'$& 0.313  $\pm$ 0.021\\
$L_X \rm(km)$& 60.8  \\
$L_Y \rm(km)$& 39.2  \\
$L_Z \rm(km)$& 27.8  \\
Surface-equivalent spherical diam. $\rm(km)$& 41.0  $\pm$ 0.7\\
Photometric surface $\rm(km^2)$ & 5288  $\pm$ 180\\
Volume $\rm(km^3)$ & $\leq 3.0\times10^4$\\
\enddata
\tablecomments{Please refer to the text for the definition of the respective quantities.}
\end{deluxetable*}

\subsection{Sphere-integrated phase curve} \label{subsec:phasecurve}
\begin{figure}
\plotone{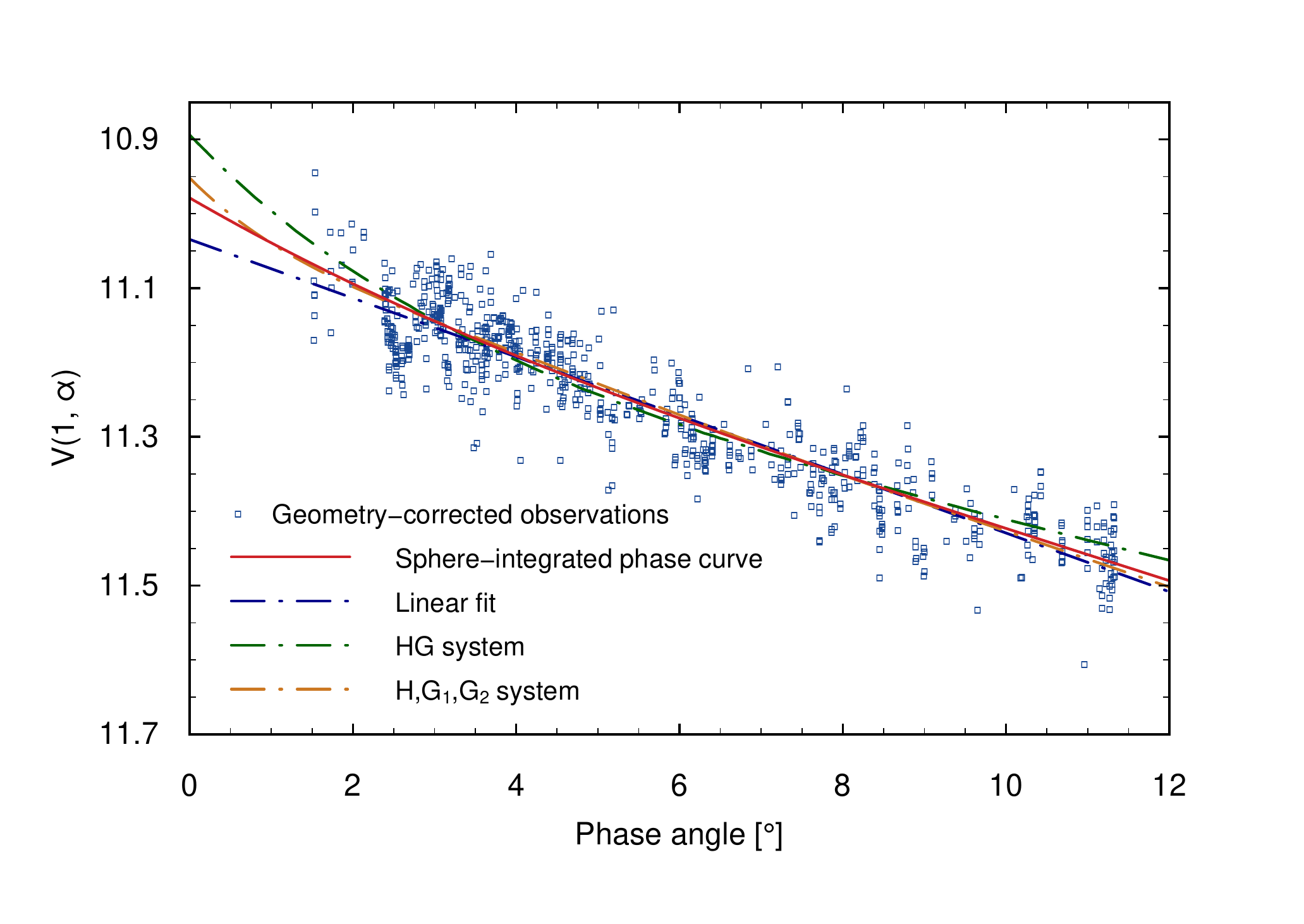}
\caption{Sphere-integrated phase curve obtained by integrating the best-fit LSL photometric function over a sphere
(red solid curve). The blue squares represent the single photometric points for which absolute calibrations and
transformations to the Johnson V band were available. Those data points have been corrected 
by multiplying the intensity of the original photometric measurement by the ratio of intensity of the spherical 
model to that of the best-fit convex shape model.
In this way the effects of changing viewing and observing geometry, as well as the rotational variations are removed.
The remaining scatter in the data is mostly due to the SNR of the measurements and to the uncertainty
in the zero points of the absolute calibrations. A fit to the data by using the IAU HG system\added{, the $\rm{HG_1G_2}$ system,} and a linear 
phase function are also shown for comparison.\label{fig:phasecurve-fig}}
\end{figure}

The disk-integrated phase curve of an object condenses the often complex parameter
space of a photometric function into a two-dimensional space.
As such, the phase curve is a useful phenomenological tool to compare
and classify different objects, and to infer the presence of physical phenomena,
as e.g. coherent backscattering. A practical difficulty in deriving phase curves, however,
is that during a single apparition -- and even more so across multiple apparitions --
in addition to the phase angle also other viewing and illumination angles change 
(i.e. aspect angle and photometric obliquity), which affects the phase curves.
This effect is more pronounced the more the object deviates from a sphere. 
Having determined through modeling the surface phase function, however, 
it is possible to compute the phase curve that the object would display 
if it were a sphere, freeing it from any dependence on the shape and thus
being only the expression of the photometric properties of the surface regolith. This is achieved by integrating
the surface phase function over a sphere in a range of phase angles, as detailed in Appendix A.
This representation, which we refer to as the \textit{sphere-integrated phase curve} corresponds to the \textit{reference phase curves}
defined by \citet{Kaasalainenetal2001} in the particular case of a sphere.
Figure \ref{fig:phasecurve-fig} shows the sphere-integrated phase curve for the LSL photometric function for Leucus (red line),
corresponding to the best-fit photometric parameters derived in Section \ref{subsec:occultationfit} and listed in Table \ref{tab:results}.
For the purpose of comparison, 
\replaced{ also a linear phase fit and a fit to the IAU HG system \citep{Bowelletal1989}  are shown.}
{fits are also shown for 1) the best-fit linear phase function ($\beta=0.0395\rm mag/\degr$) 2) the IAU HG system \citep{Bowelletal1989} and 3)
the more recently adopted IAU $\rm{HG_1G_2}$ system \citep{Muinonenetal2010}}.
\added{The latter was computed with the online tool described in \citet{Penttilaetal2016}}.

As already apparent during the 2016 apparition \citep{Buieetal2018}, and as observed for several other Trojans (see e.g. \citealt{Shevchenkoetal2012}), Leucus has a 
very subtle -- if at all -- opposition effect.
Within the observed phase angle range, 
\replaced{both the LSL phase curve and a linear phase curve with $\beta=0.0395\rm mag/\degr$ provide a good fit to the data.
The HG system less so, showing systematic variations both at the small and at the large end of the phase angle range.}
{all of the phase curves except for the HG curve -- that shows systematic variations both at the small and at the large end of the phase angle range --
provide a good fit to the data}.
\replaced
{The extrapolation at zero phase produces $H_{\rm V}$ values that differ by about 0.05 mag for the 
linear fit ($H_{\rm V-lin}$) and the LSL solution ($H_{\rm V-LSL}$).} 
{The extrapolation at zero phase produces $H_{\rm V-lin}$ and $H_{\rm V-HG_1G_2}$ values that differ from 
the LSL solution ($H_{\rm V-LSL}$) by about 0.05 mag and 0.03 mag, respectively.}
\replaced
{This is an expression of}
{These small deviations are partly due to}
  the fact that no calibrated data were available
in the V band below the phase angle of about 1.6$\degr$ to constrain the fit. \citet{Buieetal2018} did observe Leucus in the r' band at phase angles 
as low as 0.125$\degr$ during the 2016 apparition. However, there appear to be calibration inconsistencies 
between the 2016 observations and those acquired at the LCOGT in 2017 that are not fully understood. 
For this reason the 2016 observations were used as a relative data set \added{and do not contribute to our phase curve}.  
 
\added{It has long been realized (see e.g. \citealt{Oszkiewiczetal2012,Shevchenkoetal2016} and references therein) that a 
correlation exists between an asteroid's taxonomic class and the shape of its phase curve. The abovementioned tool 
by \citet{Penttilaetal2016} also performs an unsupervised taxonomic classification based on the
derived H$\rm{G_1}$,$\rm{G_2}$ solution. Interestingly, the software classifies Leucus as a D type, solely based on its phase curve.
This further supports the tentative classification by \citet{Levison2016} that was based on spectral information}.

It is also important to recall that since Leucus has a considerably larger equatorial cross section than the polar one,
oppositions with pole-on aspect would \replaced{appear}{result in measured phase curves that are} brighter 
than the \replaced{spherical model}{sphere-integrated phase curve} and conversely, 
apparitions with equator-on aspect would appear fainter than the spherical model. Given the low obliquity of 
Leucus' spin axis, however, all apparitions tend to be at near-equatorial aspect, as observed from Earth, 
and therefore Leucus never exposes its largest cross section to the observer. 
\added{This effect must be considered, e.g.,  when computing spherical equivalent diameters from observations.}

\subsection{Albedo} \label{subsec:albedo}
The geometric albedo is quite an elusive quantity to measure. 
First, it is defined for an observation geometry (0$\degr$ solar phase angle) that is rarely
observable from Earth, and in which the photometric behavior of different planetary
surfaces can wildly vary. Second, it is the result of an indirect measurement that requires 
the brightness at zero phase and a further measurement as a thermal flux or a 
geometric cross section, which adds to the total error budget. Third, the albedo 
depends in principle on the shape of the object, although this issue is less of a problem for 
dark objects as the Trojans. 
The error on the brightness at zero phase directly translates into the same relative error 
for the geometric albedo (i.e. a 10$\%$ error in the brightness would cause a 10$\%$ error in the albedo). 
In our case the geometric albedo is derived through the simultaneous fit of the photometric lightcurves
and the occultation data and by applying Eq. A3, which results in \replaced{a geometric albedo}{an accurate geometric albedo} 
determination $~p_{\rm V}=0.043 \pm 0.002$.\\

\citet{Gravetal2012} report for Leucus a much higher geometric albedo of  $p_{\rm V}=0.079 \pm 0.013$
and a spherical-equivalent diameter $D=34.155 \pm 0.646$ km derived from WISE observations. 
Those values are clearly incompatible with the occultation footprints.
Part of the reason for their overestimation
of the albedo is that they used an inaccurate $H_{\rm V}=10.70$, retrieved from the Minor Planet Center database.
Very often, those photometric measurements are acquired in non-standard photometric systems, are subject to 
inaccurate calibrations, and are therefore affected by large 
uncertainties. A further reason for the discrepancy is that 
the WISE observations happen to have occurred in the vicinity of the lightcurve minimum,
thereby underestimating the average thermal flux of Leucus.
If we use our value $H_{\rm V-LSL}=10.979$ to correct their determination by using the method proposed by 
\citet{Harris&Harris1997}, and account for the apparent visible cross-section at the time of the WISE observations,
we obtain a corrected geometric albedo $p_{\rm V}=0.048\pm0.014$ and a diameter $D=39\pm1$ km. 
With these corrections applied, the WISE determinations are compatible with our own, within the respective uncertainties.
\\

\begin{figure}
\plotone{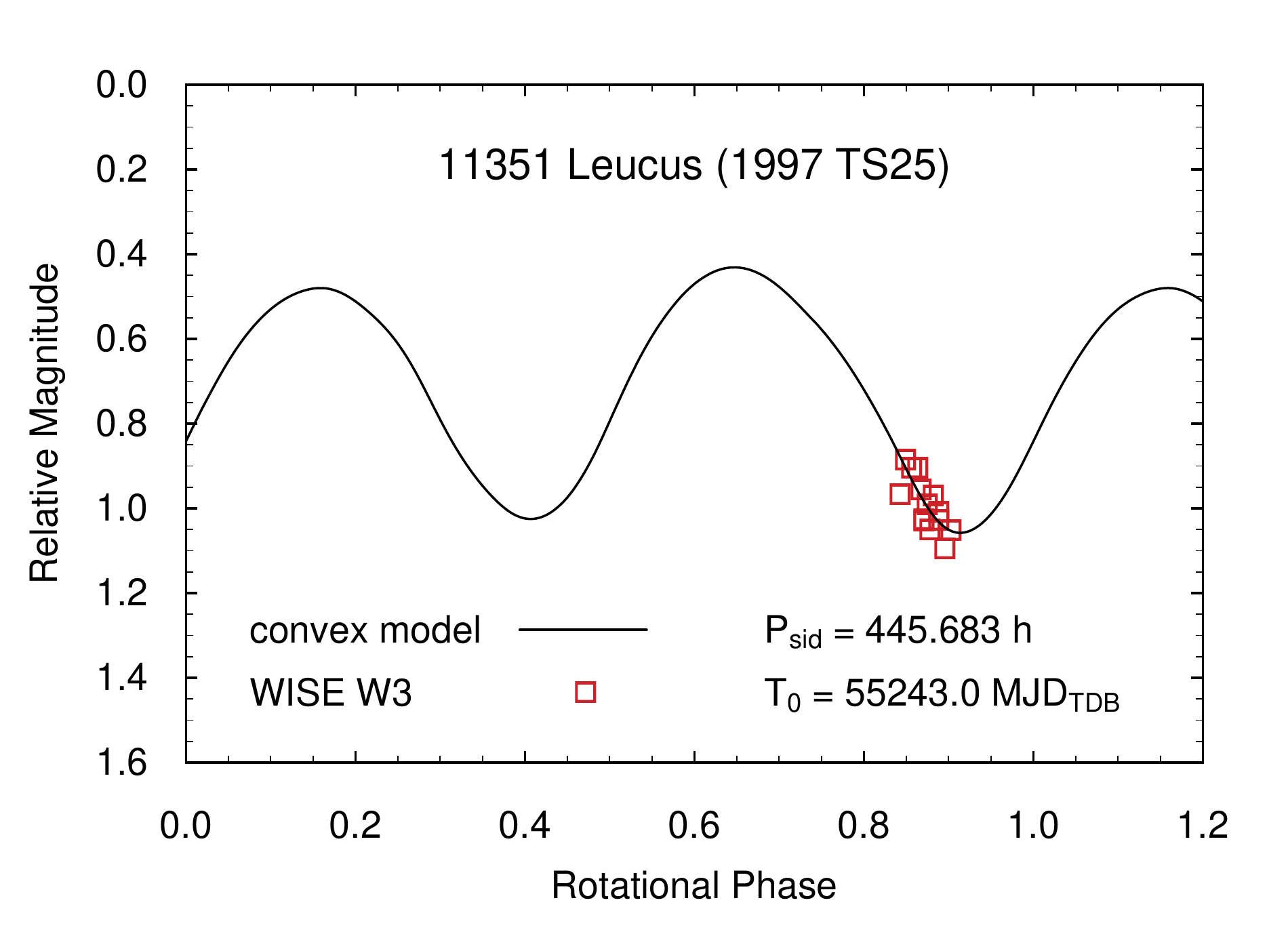}
\caption{WISE observations \citep{Gravetal2012} in the W3 12 $\mu$m thermal band are phased with a synthetic lightcurve from our convex model 
for the same epoch. Data points beyond rotational phase 1.0 are repeated for clarity. The magnitude scale for the WISE observations 
is arbitrarily offset to provide the best fit to the model curve. The plot shows that the WISE observations of Leucus were acquired near the lightcurve minimum.}
\end{figure}

The IRAS albedo and size determinations by \citet{Tedescoetal2004} (0.063$\pm{0.014}$ and 42.2$\pm{4.0}$ km, respectively)
were based on an inaccurate H-value of 10.5. By using our $H_{\rm V-LSL}$ value we update
their determinations to $p_{\rm V}=0.041 \pm 0.014$ and $D=41.9 \pm 4.0$, which 
are also in agreement with our own determinations.\\

\citet{Buieetal2020} determined geometric albedo values for the four occultation events in 2018 and 2019
by estimating the object cross-sectional area from best-fit ellipses and by using the absolute photometry reported in this paper.
They derived geometric albedo values ranging from 0.035 to 0.043 for the different occultation events, with the scatter of the measurements probably reflecting the
uncertainty in the different elliptical approximations of the occultation profiles.

\section{Discussion} \label{sec:cite}

The combination of time-resolved, disk-integrated photometry and stellar occultations is a powerful technique that
allows accurate characterization from the ground of otherwise unresolved targets.
We have determined a convex shape model that is compatible with the available occultation
footprints and, thanks to accurate absolute photometry, produces precise size and albedo estimates of Leucus.
Our model also allows us to understand and correct previous incompatible radiometric albedo and size determinations.\\
    
The accuracy of our rotation model is such that the
1$\sigma$ uncertainty on the rotation phase will be smaller than 2$\degr$ at the time of the June 2028 Lucy encounter with Leucus.
At the time of the fly-by the sub-solar latitude will be about -9$\degr$ and the South pole will be permanently 
illuminated -- although at grazing incidence -- whereas the North pole will be in its winter night.
Unfortunately, due to the slow rotation of the body, Lucy will be able to observe at most 60$\%$ 
of the surface in the 40 hours during which the object will be resolved with more than 40 pixels.
For this reason, an accurate ground-based shape model is very valuable to the mission. On the one hand 
it enables careful planning of the acquisition sequences, in order to guarantee optimum sampling.
On the other hand, it complements the data from the mission to complete the uncharted hemisphere,
similarly to what was done, e.g., in the case of the Rosetta fly-by of Lutetia \citep{Carryetal2010, Preuskeretal2012}. 
The latter is of crucial importance for the estimation of the volume of the object and hence of its bulk density.
Our convex model already serves this purpose well and represents a good second-order approximation of the shape
-- the first order being an ellipsoid \citep{Buieetal2020}. \\

With an angular size of the longest axis of 15 mas at most, Leucus 
represents a challenging target to resolve for ground-based adaptive optics, James 
Webb Space Telescope or ESO's ALMA observations. Further improvement of the shape model
in the near future can likely only come from dense stellar occultation data at further geometries, and from more, accurate absolute lightcurves.
These data could be possibly used to produce a realistic non-convex model, provided the observation geometries are favorable. 
In this respect it is important to recall that not all concavities are directly resolvable from stellar occultations. A Star-Wars Death-Star
shape, for example, would always project a convex occultation silhouette, with its concavity only being hinted at by a flat side of the contour.   \\
 
Our photometric modeling of Leucus confirms that the object is very dark and lacks a pronounced opposition effect.
These properties put Leucus in the context of other Trojan asteroids, and of other primitive, Outer-Belt objects.
Due to the limited phase angle range achievable from Earth, however, we caution from extrapolating the derived phase curve
to predict the brightness of Leucus at the large phase angles occurring on Lucy approach ($>$ 90$\degr$), because it could be in error by a considerable factor.
For this purpose it will be important to benefit \replaced{form}{from} Lucy's vantage point during the cruise phase to extend the coverage of the
phase curve to larger angles. Such measurements would also allow the use of more sophisticated photometric models
and to uniquely retrieve their parameters. Further, they would allow a reliable 
determination of the phase integral, which,  together with the geometric albedo, is critical to establish the thermal balance of the body.\\

\added{Leucus exhibits an exceptionally slow rotation, the cause of which is currently not known. 
As of today, only about 0.8\% of the about 32600 asteroids in the Asteroid Lightcurve Data Base (LCDB)
\citep{Warneretal2009} for which rotation period estimates are available, have a slower spin than Leucus.
Objects with such slow rotation cannot be explained as just belonging to the tail of a
single population of rotators \citep{Harris2002}, and some mechanism must have been
in place that slowed down their rotation. Radiation recoil forces as the YORP effect \citep{Rubincam2000} cannot explain the
slow rotation of Leucus because of its large size and heliocentric distance. More realistic possibilities
are 1) rotation angular momentum loss due to the evolution and eventual separation of a binary system with
an elongated primary \citep{Harris2002} and 2) spin-down due to reaction forces resulting from
sublimation of volatiles. \\ 
}

\replaced{\citet{Harris1994}}{\citet{Pravecetal2014}} revised the damping time scales for excited rotation as a function of asteroid size
and rotation period. 
\replaced{Taken at face value, his Figure 1 would translate into a relaxation time 
for Leucus of $\approx$ 10 Gyr -- longer than the age of the Solar System. 
In fact, given that the bulk density of Leucus is probably closer to 1 g cm$^{-3}$ than to
2 g cm$^{-3}$ -- as assumed by Harris for his estimate -- the relaxation time should be even longer.
In other words,
if excited rotation was ever present for Leucus -- either of primordial origin, or induced by
a catastrophic event -- it should be still in place as of today. }
{By assuming a bulk density for Leucus of $1\times10^3$ kg m$^{-3}$, their estimate 
would translate into a relaxation time for Leucus in the range 2.3 - 3.0 Gyr.
If excited rotation was ever present for Leucus it could be still in place as of today.}
Given the good fit of 
the lightcurves to our simple-rotation model, however, we conclude that if
an excited rotation is present, its precession amplitude must be small. Due to the short duration of the fly-by,
it is unlikely that any degree of precession can be detected by Lucy from resolved imagery.  
Instead, unresolved photometric sequences acquired by Lucy during the last few months of approach could be
used to search for multiple periodicities in the lightcurves. The detection of a non-principal rotation state would 
place additional constraints on the dynamics and the internal structure of Leucus.
 
\appendix

\section{Photometric quantities}
 For the convex shape inversion we use a photometric function defined as the product
 of a Lommel-Seeliger-Lambert disk function and a 3-parameter linear-exponential surface phase function
 \citep{Kaasalainenetal2001, Schroederetal2013}.
The corresponding \textit{radiance factor} ($I/F$) can be written as:

 \begin{equation} 
 I/F= A_{LSL}~\mu_{0} \left(\frac{1}{\mu+\mu_{0}} +c\right) f(\alpha)
   \end{equation} 

where $A_{LSL}$ is the Lommel-Seeliger Lambert albedo, $\mu_0$ and $\mu$ are the incidence and emission angles, respectively,
and $c$ is the weight for the Lambert contribution. The term $c$ can be either constant, or a function of the 
phase angle $\alpha$. In the latter case the Lommel-Seeliger Lambert disk function is equivalent to the Lunar-Lambert 
disk function in the formulation of \citet{McEwen1996}.
The term $f$($\alpha$) is the surface phase function (expressed in intensity), which, following \citet{Kaasalainenetal2001} we choose to be of the form : 

 \begin{equation} 
f(\alpha) = A_0~e^{-\alpha/D} + k\alpha +1
   \end{equation} 

where $A_0$ and $D$ are parameters that determine the amplitude and angular width of the exponential term, respectively,
and $k$ is the slope of the linear component. The phase angle $\alpha$ is expressed in radians.
With this formalism the geometric albedo for a sphere is:

 \begin{equation} 
p_{V} = A_{LSL}~\left(\frac{1}{2}+\frac{2}{3}c_0\right) (A_0+1)
   \end{equation} 
where $c_0$ is the value of the Lambert weight at zero phase angle.  

The disk-integrated phase function for a sphere,  expressed in intensity and normalized to unity at zero phase is \citep{Lietal2015, Lietal2020}:

 \begin{equation} 
\Phi_{LSL}(\alpha) = \frac{[4c/(3\pi)~(sin(\alpha)+(\pi-\alpha)~cos(\alpha))+(1+sin(\alpha/2)~ln(tan(\alpha/4))~tan(\alpha/2))]}{(4c_0/3 +1)}~ \frac{f(\alpha)}{(A_0+1)}.
   \end{equation}   

\acknowledgments
Research at the DLR was funded by the \textit{DLR Programmatik Raumfahrtforschung und-technologie} through the grant 2474029 \textit{Lucy}.
Part of this work was supported by the Lucy Mission which is funded through the NASA Discovery program on contract number NNM16AA08C.\\

%





\bibliography{Leucus_Mottola_etal}{}
\bibliographystyle{aasjournal}



\end{document}